\documentstyle[12pt]{article} 
\input epsf 
\begin{document} 
\title{  On The Possible Detection Of Massive
Stable Exotic  Particles At The LHC } 
\author{ Aleandro Nisati, 
Silvano Petrarca and Giorgio Salvini \\[1.5em]
Dipartimento di Fisica,\\ Universit\`a di Roma {\it La Sapienza},
\\ and INFN, Sezione di Roma {\it La
Sapienza}\\ P.le A. Moro 2, 00185 Roma, Italy\\[0.5em] 
%\bigskip 
%\\[1.5em] 
% PRELIMINARY VERSION 
%\\[0.5em]
}
\maketitle \bigskip
 
\begin{abstract}
The possible detection of massive quasi-stable exotic particles 
at the high luminosity hadronic colliders is discussed. In the 
coming ten years the LHC, now under preparation, has the 
best opportunity to observe them at the TeV scale.
 The present
design of the ATLAS detector, that has been almost irreversibly 
decided, may turn out to be flexible enough to allow the 
detection of this interesting class of exotic particles. The 
trigger acceptance, the track reconstruction and the particle 
identification are studied. The necessity of a good measurement of 
the ionization loss in the muon sector of the detectors is 
recommended.
\end{abstract}
\bigskip
{\leftline{ROME1-1177/97}}
\vfill
\newpage
\section{Introduction}
\par

 In the next years the Large Hadron Collider (LHC) will be 
 the main instrument potentially
 capable of producing new particles  
 beyond the Standard Model over a wide
 mass range around the TeV scale.

 There are a lot of possible extensions of the Standard Model,
  like SUSY, compositeness,
 grand unification, left-right symmetric models etc., that foresee 
 a specific enlargement of the known particle spectrum with exotic 
 not yet observed states.
  The physics for LHC beyond the Standard Model has been extensively
 studied at the LHC workshop
 (Aachen  1990)\cite{LHC} and there is also 
 an excellent review of this subject in the proceedings of the workshop 
 for the SSC (Berkeley 1987)\cite{SSC}. 

  In this article we consider an interesting, wide-ranging, 
  class of exotic particles that are
  characterized by a few model independent common 
 features:
%\cite{Errede,Tata}:

a)    they have large masses and are produced with $\beta < 1$
      at the collider;

b)    they are stable or quasi-stable, i.e. their life-time must be larger
	than

	\quad $10^{-7}$ s, so that they  decay  outside the
	detector; 

c)    they have electric charge either integer or fractional
	  multiple of the 

 \quad proton charge.

 Actually, several theories beyond the Standard Model foresee 
 charged particles, both hadrons and leptons, with
 large masses and, althought many exotic particles are
 unstable, often the lowest (or the next-to-the lowest) 
 lying state may be stable or
 quasi-stable\cite{Errede}\cite{kaori}.
 Throughout this article 
 we call them Massive Stable Exotic (MSE) particles.
 
 Here we briefly review some of the results of the most recent
 searches based on this philosophy.
  One example of MSE is given by heavy quarks belonging 
 to  higher representations of $SU(3)_C$\cite{Ma_ecc}.
  Recently a color sextet of quarks $Q$ has been 
 proposed 
 in order to explain the dynamical breaking of the
 electroweak symmetry \cite{Watanabe}.
 The quark sextet, taking the role of a condensate,
 forms a doublet of weak $SU(2)_{L}$
 (U D) with mass range: $300 < m_U, m_D <400$ GeV/$c^2$. 
  The lightest state of these 
 quarks may be stable or it may decay into normal quark and 
 long-lived leptons 
 after being hadronized in an exotic colorless hadron.

The cross-section of this exotic reaction, which has been 
calculated in the gluon fusion model in ref. \cite{Watanabe},
turns out to be 
around one order of magnitude greater than the top quark production 
cross-section because of the larger color factors of the sextet. 

 The CDF collaboration at the TEVATRON 
 has published several analyses
 setting upper limits on the MSE particles cross-sections,
 see ref. \cite{CDF} and references therein. 
 The theoretical cross-sections have been used by CDF
 to set bounds on the mass of fermionic color triplets,
 sextets, etc. In particular,
 at the Warsaw Conference\cite{CDF},
 CDF has presented
 a search for
 particles with low velocity,
 $\beta\gamma < 0.65$, with $p_T > 30$ GeV/c
 and $dE/dx$ $\geq$ 1.8 (2.5)
 times larger than expected
 for a particle at minimum in 
 the central tracker (silicon tracker).
 The non-observation of MSE particles
 is used preliminarly  to set a lower limit
 of $190$ GeV/${\rm c}^2$
 on the mass
 of a stable color triplet quark.
 
 Other examples of MSE particles come from SUSY.
 In ref.\cite{Tata}  the production of
 MSE particles 
 at the hadronic colliders has been examined
 and also the production cross-sections for
 different types of SUSY MSE particles 
 together with a detailed study of
 their
 experimental signatures has been reported.
  For these channels, the value of the cross-sections 
  are lower than the top 
 quark pair cross-section but there is still a large range 
 of the parameters
 for which the reaction is potentially visible at LHC.
 
 In some realizations
 of SUSY, there can be charged, 
 long-lived 
 supersymmetric particles like charginos or staus
 that behave like MSE particles.
 
 At LEP1, searches for MSE particles were performed\cite{LEP1}  
  obtaining the  limit (ALEPH)   on
 the production cross-section of $\sim 1.5$ pb at $95\%$ CL
 in the mass range
 $34 \div 44$ GeV/${\rm c}^2$.
 At LEP2, both ALEPH\cite{al} and DELPHI\cite{de}
 have searched for MSE particles up to ${\sqrt s}=172$ GeV.
 DELPHI gives upper limits at $95\%$ CL  on 
 the  cross-section of pair-produced
 MSE particles 
 in the range of 
 $0.4 \div 2.3$ pb for masses from 45 to 84 GeV/${\rm c}^2$
 and electric charge $\pm e$ and $\pm 2/3 e$. ALEPH gives a model independent 
 at $95\%$ CL upper limit on the production cross-section of $0.2 \div 0.4$
 pb for masses between $45$ and $86$ GeV/${\rm c}^2$.
 All these 
 (unsuccessful) searches could only establish the upper limits we 
 reported.
 
 These results imply, when converted in limits on SUSY particles, 
 that long-lived charginos are excluded for masses below 86 GeV/${\rm c}^2$ and
 long-lived left-handed (right-handed) smuons and staus below 
 $68 (65)$ GeV/${\rm c}^2$.

  The energy loss in matter of an MSE particle
  has been carefully studied 
 in ref.\cite{Tata} (see also ref.\cite{Errede}). As far as the energy 
 loss for ionization is concerned, for $\beta > {1\over2}$ and for particles 
 with $M > $100 GeV/$c^2$ the range in iron exceeds several meters
 and increases with $M$.  
 Moreover,
 for strongly interacting MSE particles 
 the penetration length in matter does not 
 change significantly
 for $\beta \leq 0.7$ and $M= $100 GeV/${\rm c}^2$\cite{Tata}. 
 
  Nevertheless, it is important to note that an MSE particle, 
 because of its low $\beta$ value, 
 produces an anomalous energy loss for ionization in matter that can be
 used as a distinctive feature for its identification.

 This discussion supports the statement that an MSE behaves like
 a massive muon  with a velocity 
 considerably lower than $c$ and consequently
 with high energy losses in matter.
  Therefore the selection and 
 identification of an MSE particle 
 may naturally be done measuring the time of 
 flight, the momentum and the ionization of a particle 
 that is able to escape from 
  hadronic calorimeter.  
 Following these lines two of us  have
 already discussed, in  a previous note\cite{PetSal}, the problems 
 connected with
 the detection of MSE particles at 
 LHC with the air-core apparatus ATLAS. In particular in that paper
 it has been suggested to add to the apparatus a specialized "window"
 devoted to the observation of MSE particles. 
 
 At the present stage  of
 the design of the LHC detectors, the complexity of the structures
 makes it unthinkable to add specific sectors inside the detectors 
 and anyhow  we  leave these windows to the rather distant
 future.
  Here 
 and in the rest of this article we focus the discussion on the 
 case of the ATLAS detector.
 For the sake of completeness we 
 have shown in Figs. 1 the present design of the ATLAS 
 detector\cite{techprop}.

 Nevertheless, we believe that it is important and still 
 possible to maintain in LHC the capacity to search for exotic 
 particles like MSE,
 undertaking some final decisions on the MDT 
 detector and
 using suitable off-line algorithms for the event reconstruction.
 
	In this paper we analyse the present design of ATLAS in 
 order to verify how its structures could allow the detection of 
 MSE particles taking into account the specific qualities we have 
 stated before, and without altering the hardware design. As we 
 have already stated, this is the main point of this note.

\par
\bigskip
\section{\bf Trigger Acceptance of MSE Particles}

 First of all we have to check  the possibility that an MSE particle
 can be accepted by the trigger. At LHC the bunch crossing rate
 of $25$ ns  and the high luminosity set stringent limits on the trigger.

	The ATLAS trigger\cite{techprop}
 is organized in three sequential levels 
which must reduce the initial input rate of 40 MHz to the first 
level trigger progressively to a rate of $\sim$ 100 Hz; that is the 
maximum input rate of the data storage system and corresponds to 
an acquisition of $\sim$ 100 MB/s.

	In the following we will ascertain the possibility that an
MSE particle is accepted by the first level trigger that is the 
most rapid and almost completely hard-wired. We consider the 
implications of the higher trigger levels a less severe problem, 
which can be addressed in the future.

Assuming that 	an MSE particle is similar to  massive muon,
 we now study the 
possibility that 
the first level muon trigger may allow the 
detection of an MSE particle. 

 In ATLAS, the muon trigger  is composed by 3 stations
 each made of 2
 RPCs (Resistive Plate Chambers) read out in the  
 2 orthogonal views, located at 
 $\sim 7.0 \sim 7.5 \sim 10.5$ m (we call them station 1, 2, 3
 respectively)
 from the beam line. 
  The RPCs are the most rapid detectors of the apparatus: they 
 have a typical gaussian time resolution $\sigma = 1.5$ ns.
 Taking into account the signal delay along the collecting strips,
 the muon trigger system assigns a temporal gate that is,
 in one of the possible choices,  of 18 ns
 to each RPC in order to allow an efficient temporal coincidence of
 the innermost and the outermost trigger stations to the crossing
 of a high energy muon.  
 After that, the event is stored in  pipe-line memories, inside the 
 apparatus, waiting for the decision of the first level trigger.
  If the track  gets a valid identification, the event is taken
  and the data go to the second level trigger, 
  otherwise the event is lost.
 The decision 
 to acquire the data from the pipe-line must be taken
 in a maximum time of $2.5 \mu$s that is the latency time 
 of the first level trigger.
 
 The RPCs (and all the other part of the apparatus)
 are locally synchronized with the LHC clock of $25 n$s
 taking into account the relativistic
 delay among 
 different parts due to the relative distance.
 This fact is a potential source of rejection 
 of an MSE particle with $\beta < 1$.
  However the temporal gate of 18 ns provides a window large enough for 
 the detection of MSE particles
 as we will show.

 We distinguish two muon trigger schemes: one is the same trigger 
 proposed for high luminosity ($\sim 10^{34} cm^{-2} s^{-1}$) 
 and high $p_T$ ($\geq 20 $ GeV/c) muons in which 
 all the three stations of RPC
 are in coincidence.
 The second is the trigger proposed for low $p_T$  
  ($\geq 6$ GeV/c) muons and low
 luminosity  ($\sim 10^{33} cm^{-2} s^{-1}$)
 in which only the two inner stations
 (1 and 2) are in coincidence (see Fig. 1).
 
 We call these two trigger schemes trigger A and trigger B respectively.
 In both cases the first level trigger accepts the event when
 a coincidence is obtained among 
 the three (or two in the scheme B) 
 temporal windows of 18 ns. In this case the data taken from   
 the RPCs are labelled by a bunch crossing number which  identifies
 the event, this number is the same for the data of the same event
 coming from 
 the other  detectors of the apparatus. 

 In Fig. 2 we give an estimate, based on the
 parameters explained above, of the trigger efficiency
 for schemes A and B,
 normalized to  the trigger efficiency for an
 ultra-relativistic muon, as a function of $\beta$ for
 particles coming out of the inner detector.
 We see that, for the trigger A, 
 at $\beta = 0.5$ we still have a good efficiency ($\sim 50 \%$)
 that decreases rapidly and approaches
 0 at $\beta \sim 0.35$.
	This is due to the fact that the trigger does not require 
the bunch crossing time as a further coincidence. The main 
coincidence is then between station 1 and 3 that are at a 
distance of $\sim 3$ m; in this case 18 ns is a large enough time to 
accept MSE particles with $\beta\geq  0.5$. 

	For the sake of completeness in Figs. 3A and 3B we have 
shown the main features of an  MSE particle coming out from the 
central calorimeter with velocity $\beta c$.  In Fig. 3A we have 
reported the delay (ns/m) of an MSE particle with respect to an 
ultra-relativistic particle, and in Fig. 3B we have plotted the 
energy loss rate normalized to its minimum value. 
	In the case of trigger B, because of the small distance 
($\sim 0.5$ m) between the two RPC stations, the efficiency remains 
high over a larger region, and goes down at a very low $\beta$ value:
$\beta \sim 0.15$.
 We must note that trigger B can be used only when LHC runs 
at low luminosity, because it does not guarantee the necessary 
background rejection in the case of high luminosity.

	On the other hand, due to the very fast increase of the 
energy loss with decreasing $\beta$ values, the MSE particle may stop 
inside the hadronic calorimeter. In Fig. 4 we have shown the 
range of an MSE particle in iron as a function of $\beta$ for
$ M = 0.2, 0.6$ and 1 TeV/$c^2$. Considering that the inner
 calorimeter system 
corresponds to about 2 m of iron, we see that only MSE particles 
with masses $M \geq 1$ TeV/$c^2$ could escape off the calorimeter when 
$\beta \leq 0.25$.
	An MSE particle stopped inside the calorimeter is typically 
lost because the kinetic energy deposited into the calorimeter 
is, in almost every case, below the threshold of the hadron 
calorimeter trigger.
In any case the calorimeter system will
contribute to the MSE identification with the
measurements of
the energy deposition, particularly in the more external
layer.

\bigskip
\section{MSE Particle Identification}

 The track identification of an MSE particle  escaping 
 the hadronic calorimeter in the high $p_T$ muon region 
 is the natural task of 
 the ATLAS specialized structure
 of precision chambers, composed of MDTs (Monitored Drift Tubes), which 
 are dedicated to
 determining, with a good precision, the position of a 
 charge particle. 

 We recall that in this region an 
 air-core toroidal superconducting magnet generates a magnetic field 
 of about 0.5 T to allow the measurements of particle momentum.
 The precision chambers are composed
 by two multilayers, and each of them is composed
 of three (four in the inner station) MDT layers.
 A typical track goes through about 
 twenty MDTs before escaping from the apparatus.
  A 
 single MDT is a cylindrical counter of 3 cm diameter; it is 
 basically filled with Argon at 3 atmosphere absolute pressure. It 
 has a typical spatial resolution $\sigma = 80$ $ \mu$m and a maximum drift 
 time of 500 ns. The resulting spatial resolution for the single 
 particle is about 30 $\mu$m\cite{frontend}. 
 
 When an event has been accepted by the first
 level trigger, all the data coming 
 from the MDTs for the following $500 \,n$s 
 are labelled with the bunch crossing
 number and are extracted from the pipe-line memories. The track recontruction
 for an MSE particle will be done by taking into account in the hit analysis 
 the delay along the track  due to $\beta < 1$.
 Note that there is the possibility (depending on
 the position of the crossing point respect to the MDT wire)
 that the more delayed hits
 could be associated with the following bunch crossing number.
 This is a problem that must be faced  in a more complete analysis.
 Here
 we want to note that due to the large MDT
 drift time ($500 \,n$s) this effect should be
 very small.

 	In order to estimate the mass measurement accuracy of an MSE 
particle, it is necessary to know the error in the measurement of 
the $\beta$ value
that is connected to the MDT intrinsic space resolution and to 
the maximum drift time.

	By combining the time measurements available from MDTs and 
RPCs along the particle trajectory we get a conservative estimate 
for the $\beta$ accuracy: 
$${\Delta\beta\over\beta}\simeq 0.1 \cdot \beta \; .$$

	The measurement of the mass M of the MSE particle is 
obtained using the relation $M = P/{\beta \gamma}$ ,
 and the error is given by
 $${\Delta M/ M} = \sqrt{(\Delta p/p)^2 +(\Delta \beta \gamma/{\beta
\gamma})^2}\, .$$

 The term  $\Delta \beta \gamma/{\beta\gamma}$
satisfies the relation 
$$ {\Delta \beta\gamma \over{\beta \gamma}}
= \gamma^2 {\Delta\beta\over{\beta}} \simeq
0.1 \cdot \gamma^2 \beta$$
and dominates, over a large range of $\beta$, 
 the term $\Delta p/p \simeq 0.025 \div 0.10$
for the ATLAS muon spectrometer\cite{techprop}. Therefore we conclude 
that the MSE mass can be measured with a resolution of the order 
of 10\%.

	Besides the strong signature of the delay needed to 
reconstruct the track, a MSE particle is identified by the energy 
loss for ionization that can be much higher than that of muon at 
minimum. We see, from
Fig. 3B, that already at $\beta \sim 0.6$ 
the energy loss is around twice the minimum and of course it 
increases rapidly for a
lower value of $\beta$. Already with a specific 
ionization of twice the minimum, the deposit of primary charge 
into the twenty MDTs
through which
the MSE particle passed will 
be statistically recognizable. Therefore the signature of high 
energy loss is a powerful physical quantity that must be 
considered to make a clear identification of an MSE particle. 

	Up to now, the MDT front-end electronics is not completely 
settled. A feasibility study to achieve a measurement of charge 
in the MDTs front end equipment is under way\cite{frontend}. We know that 
the economical convenience to measure the charge in each channel 
is still under discussion. This measurement is important from the 
general point of view to distinguish real muon tracks from the 
background of neutron and gamma hits. We underline here that the 
measurement of $dE/dx$ in the MDTs is also fundamental to any 
future hope to observe MSE particles. 

\bigskip
\section{Conclusion}

We believe that there are excellent reasons for looking 
for MSE particles. It is important that this search be done at the
LHC accelerator, which will keep the energy and luminosity 
supremacy for many years. As we said, the range of MSE particles 
with a remarkable ionization loss may be quite high.

	We have seen that the first level muon trigger allows the 
gathering of MSE particle data over a large space of the MSE 
particle parameters (mass and $\beta$). 

	For the runs at low luminosity, the muon trigger made by two 
RPC layers provides a larger window. The track reconstruction of 
an MSE particle must take into account the temporal delay of these 
particles, and this delay is the first strong signature of an MSE 
particle. The measure of the ionization losses in the MDTs which 
is, in our opinion, of great importance, will strongly improve the 
identification of MSE particles in the muon system. The inner 
detector and the hadronic calorimeter can complete the track 
reconstruction. 

	The present analysis of this problem is far from being 
complete. In fact the higher levels of the muon trigger ought to 
be studied and the track reconstruction for an MSE particle should 
be evaluated with the necessary accuracy. We think also that 
other methods could be followed in the research of MSE particles 
with the ATLAS detector.

	The aim of this note is to point out that the strategy of 
searching for MSE particles is an important field of study which can 
be investigated at the LHC with the present ATLAS detector.

\bigskip
\section{Acknowledgements}

 We are grateful to  Emilio Petrolo, Ludovico Pontecorvo, 
 David Stuart, Lucia Zanello,  Stefano Veneziano 
 for  general discussions and suggestions.

\par 

\newpage

\section{Figure Captions}

\noindent
Fig. 1:

\noindent
Side view of a quarter of the ATLAS detector showing the locations of different 
        subsystems
        (taken from ref.\cite{techprop}).
         In the innermost part of the apparatus is visible
        the inner tracker (IT); the electromagnetic and hadron calorimeter are
        indicated respectively by  EM and HC. The muon system in the barrell
        extends from  about 5 to 10 m in radius and is composed by three 
        tracking stations and three
        trigger stations  indicated respectively
        MDT and RPC  Stations 1, 2, 3. The barrell region extends over
        $ | \eta | \le 1.05 $. In the figure the layout of the forward
        muon system is also shown.
         
\bigskip

\noindent
Fig. 2: 

\noindent
Estimate of the (muon) trigger efficiency (normalized
 to the efficiency of an ultra-relativistic muon)  as a function
of the $\beta$ values of an MSE particle escaping the central calorimeter.
The curves (A) and (B) refer to trigger scheme A and B respectively.
\bigskip
        
\noindent
Fig. 3:

\noindent
A) Delay per meter of a particle as a function of $\beta$.

\noindent
B) Energy loss rate normalized to the value at the minimum as a function
  of $\beta$.
The 
 scale in
$\beta \gamma = \beta/\sqrt{1-\beta^2} $ is also shown.
\bigskip

\noindent
Fig. 4: 

\noindent
Range of an MSE particle in iron as a function of $\beta$ for three different
mass values.
Note that the central calorimeter corresponds to about 2 m of iron.
The scale in $\beta\gamma$ is also shown.


\bigskip 
\begin{thebibliography}{99}
 
\bibitem{LHC}
 Proc. LHC Workshop Aachen 4-9 October 1990 vol.II ;
 eds: G.~Jarlskog, D.~Rein.


\bibitem{SSC}
 A.~M.~Litke et al.; Proc. Workshop on experiments, detectors and
  experimental areas for the supercollider. ( Berkeley, 1987), eds.
  R.~Donaldson and M.~G.~D.~Gilchriese (World Scientific, 
  Singapore, 1988) pag. 853.


\bibitem{Errede}
 S.~Errede and S.~H.~H.~Tye, in Proc. of the 1984 summer study on the
     design and utilization of SSC, Snowmass, Colorado, 1984, eds.
     J.~Morfin and R.~Donaldson, pag. 175.
 
\bibitem{kaori}
 K.~Enqvist, K.~Mursula, M. Roos:
 Nucl. Phys. {\bf B226}, (1983) 121; 
 
 H.~E.~Haber and G.~Kane:
 Phys. Rep. {\bf C 117}, (1985) 75; 
 
 H.~Fritzsch:
 Phys. Lett. {\bf B 78}, (1978) 611;
 
 P.~Fishbane, S.~Meshkov and P.~Ramond:
 Phys. Lett. {\bf B 134}, (1984) 81;

 R.~Barbieri, Riv. Nuovo Cimento {\bf 11}, (1988) 1.

\bibitem{Ma_ecc}
 E.~Ma:
 Phys. Lett. {\bf B58}, (1975)442;

 G.~Karl:
 Phys. Rev. {\bf D14}, (1976)2374;

 F.~Wilczek and A.~Zee:
 Phys. Rev. {\bf D16}, (1977)860;

 Y.~Ng and S.~H.~H.~Tye:
 Phys. Rev. Lett. {\bf 41}, (1978)6;

 H.~Georgi and S.~Glashow:
 Nucl. Phys. {\bf B159}, (1979)29.
 
\bibitem{Watanabe}
 H.~Tanaka and I.~Watanabe:
 Int. J. Mod. Phys. {\bf A7} (1992) 2679.

\bibitem{CDF}
K.~Maeshima for the CDF Collaboration: 28th 
International Conference on High Energy Physics, Warsaw, 
Poland, 25-31 July 1996;
		
The CDF Collaboration, F. Abe et al, Phys. Rev. {\bf D 46} 
(1992) 1889;

The CDF Collaboration, F. Abe et al, Phys. Rev. Lett. 
{\bf 63} (1989) 1447.


\bibitem{Tata}
 M.~Drees and X.~Tata:
 Phys. Lett. {\bf B252}, (1990)695.


\bibitem{LEP1}
 The ALEPH Collaboration:
 Phys. Lett. {\bf B 303}, (1993) 198;


 The DELPHI Collaboration:
 Phys. Lett. {\bf B247}, (1990) 157;

 
 The OPAL Collaboration:
 Phys. Lett. {\bf B252}, (1990)290.

\bibitem{al}
 The ALEPH Collaboration:
 Search for pair-production of long-lived
 heavy charged particles in $e^ + e^-$ annihilation,
 CERN-PPE/97-041.

\bibitem{de}
 The DELPHI Collaboration:
 Search for Stable Heavy Charged Particles
 in $e^ + e^-$ Collisions at $\sqrt s$=130-136,161 and
 172 GeV,
 CERN-PPE/96-188.


\bibitem{PetSal}
 S. Petrarca, G. Salvini:
 Department of Physics University of Rome "La Sapienza",
 Internal Note n. 999, 18 September 1992;
 Search For Stable Exotic Massive Particles
 At LHC By An Instrumented
 Air-Core Toroid (Ascott Type).

\bibitem{techprop}
 ATLAS Collaboration:
" Technical Proposal for a General-Purpose pp Experiment
 at the Large Hadron Collider at CERN";
 CERN/LHCC/94-43; LHCC/P2 15 December 1994.
 
\bibitem{frontend}
 E.~Hazen, J.~Shank:
 "Status of the Front End Electronics for the MDT System"
 ATLAS Internal Note: MUON-NO-111, 1 March 1996.
\end{thebibliography}
\end{document}